\documentclass[twocolumn,aps,]{openjournal}
\usepackage{enumitem}
\usepackage{graphicx}
\usepackage{bm}
\usepackage[colorlinks=True]{hyperref}
\usepackage{natbib}
\usepackage{booktabs}
\usepackage{amsmath}
\usepackage[mathlines]{lineno}
\usepackage[table,xcdraw]{xcolor}
\begin{document}

\title{The Effect of Splashback on Weak Lensing Mass Estimates of Galaxy Clusters and Groups}

\author{Yuanyuan Zhang$^{1, 2}$, 
Susmita Adhikari$^{3}$, Matteo Costanzi$^{4, 5, 6}$, Josh Frieman$^{7, 8, 9}$, Jim Annis$^9$, Chihway Chang$^{7, 8}$}
\affiliation{
$^1$NSF's National Optical-Infrared Astronomy Research Laboratory, 950 N Cherry Ave, Tucson, AZ, USA \\
$^2$Mitchell Institute for Fundamental Physics and Astronomy, 4242 TAMU 576 University Dr, College Station, TX, USA \\
$^3$Indian Institute of Science Education and Research, Dr. Homi Bhabha Road, Paashan, Pune India \\
$^4$ Astronomy Unit, Department of Physics, University of Trieste, via Tiepolo 11, I-34131 Trieste, Italy \\
$^5$INAF-Osservatorio Astronomico di Trieste, via G. B. Tiepolo 11, I-34143 Trieste, Italy \\
$^6$Institute for Fundamental Physics of the Universe, Via Beirut 2, 34014 Trieste, Italy \\
$^7$Department of Astronomy and Astrophysics, University of Chicago, Chicago, IL 60637, USA \\
$^8$Kavli Institute for Cosmological Physics, University of Chicago, Chicago, IL 60637, USA\\
$^9$Fermi National Accelerator Laboratory, P. O. Box 500, Batavia, IL 60510, USA \\
}

%\email{$\ast$ }

\begin{abstract}
The splashback radius of a dark matter halo, which corresponds to the first apocenter radius reached by infalling matter and substructures, has been detected around galaxy clusters using a multitude of observational methods, including weak lensing measurements. In this manuscript, we present how the splashback feature in the halo density profile affects galaxy cluster masses derived through weak lensing measurements if it is not accounted for. We find that the splashback radius has an  increasingly large effect on group-sized halos towards $M_{200m} \sim 10^{13.5} \mathrm{M_\odot}$. Depending on the model and the radial scale used, the cluster/group masses can be biased low by more than 0.1 dex. This bias, in turn, would result in a slightly lower $\Omega_m$ value if propagated into a cluster cosmology analysis. 
The splashback effect with group-sized dark matter halos may become important to consider given the increasingly stringent cosmological constraints coming from optical wide-field surveys. 

\end{abstract}
%\begin{keywords}
%(cosmology:) large-scale structure of Universe -- galaxies: clusters: general
%\end{keywords}

\maketitle

\section{Introduction}

%https://scholar.google.com/scholar?hl=en&as_sdt=0%2C14&q=galaxy+cluster+lensing+splashback&btnG=

As massive dark matter halos accrete matter or substructures, a so-called splashback radius can be found at their outskirt, referring to the radius where infalling matter or substructures reach the apocenter for the first time \citep{2014JCAP...11..019A, 2014ApJ...789....1D, 2015ApJ...810...36M, Shi:2016lwp}. It has been revealed that observational studies can measure the splashback radius through a transition in the halo density distribution  \citep[the caustic feature in the density profile][]{2017ApJ...841...34M, 2017ApJ...843..140D}, which opens up additional channels to study this unique astrophysical effect.

Measuring the splashback effect of galaxy clusters -- the largest dark matter halos -- has turned out to be particularly successful. Splashback radius of galaxy clusters have been detected using cluster galaxy (corresponding to subhalos in simulations) number densities \citep{2017ApJ...841...18B, 2016ApJ...825...39M}, and/or through weak lensing measurements \citep{ 2017ApJ...836..231U,2018ApJ...864...83C, 2019MNRAS.485..408C, 2019MNRAS.487.2900S} of  matter densities, or even through stellar light distribution \citep{2021MNRAS.500.4181D, 2021MNRAS.507..963G}. With those measurable effects, splashback effect has thus brought forward additional pathways to study cosmology \citep{2018JCAP...11..033A} and dark matter properties \citep{2020JCAP...02..024B}, the cluster galaxy evolution and accretion history \citep{ 2021ApJ...923...37A, 2021ApJ...911..136B}, assembly bias \citep{2016PhRvL.116d1301M, 2017MNRAS.470.4767B, 2019MNRAS.490.4945S} as well as new definitions for halo mass and halo boundaries \citep{2015ApJ...810...36M, 2021ApJ...909..112D}. 

Because the splashback phenomenon causes a measurable effect on galaxy cluster matter densities measured through weak lensing, we may wish to turn around on this discovery and examine the need to model the splashback feature in weak lensing analyses for other scientific purposes. Especially, weak lensing measurements of galaxy cluster masses provide critical input data to cosmology analysis. These mass measurements frequently rely on models of the halo matter radial distribution \citep[e.g.,][]{2019ApJ...875...63M}, and models incorporating the splashback effect have been deployed in some weak-lensing analyses \citep{2017ApJ...836..231U}. 

In this short manuscript, we demonstrate the possible effects of  the splashback feature on galaxy cluster weak lensing measurements using simulations, covering especially the low-mass clusters/groups towards $M_{200m} \sim 10^{13.5} \mathrm{M_\odot}$. This mass range is of particular interest for the current photometric cluster surveys and upcoming cluster surveys at all wavelengths, which can probe low mass and group scale halos. We consider a cluster lensing model that has been used by some previous optical surveys like SDSS \citep{2014MNRAS.439.1628Z} and DES \citep{2019MNRAS.482.1352M} as well as a model that is used to model splashback effect \citep{2014ApJ...789....1D}. 
The rest of this paper is organized as the following: In Section~\ref{sec:sims}, we introduce the measurements on simulations, and Section~\ref{sec:model} visually demonstrates the differences between the two models. Section~\ref{sec:cosmo} discusses the cosmological effect, and Section~\ref{sec:concl} summarizes the results.

\section{simulation}\label{sec:sims}

\begin{figure*}
	%\includegraphics[width=\columnwidth]{measurement_profile.png}
	\includegraphics[width=2.1\columnwidth]{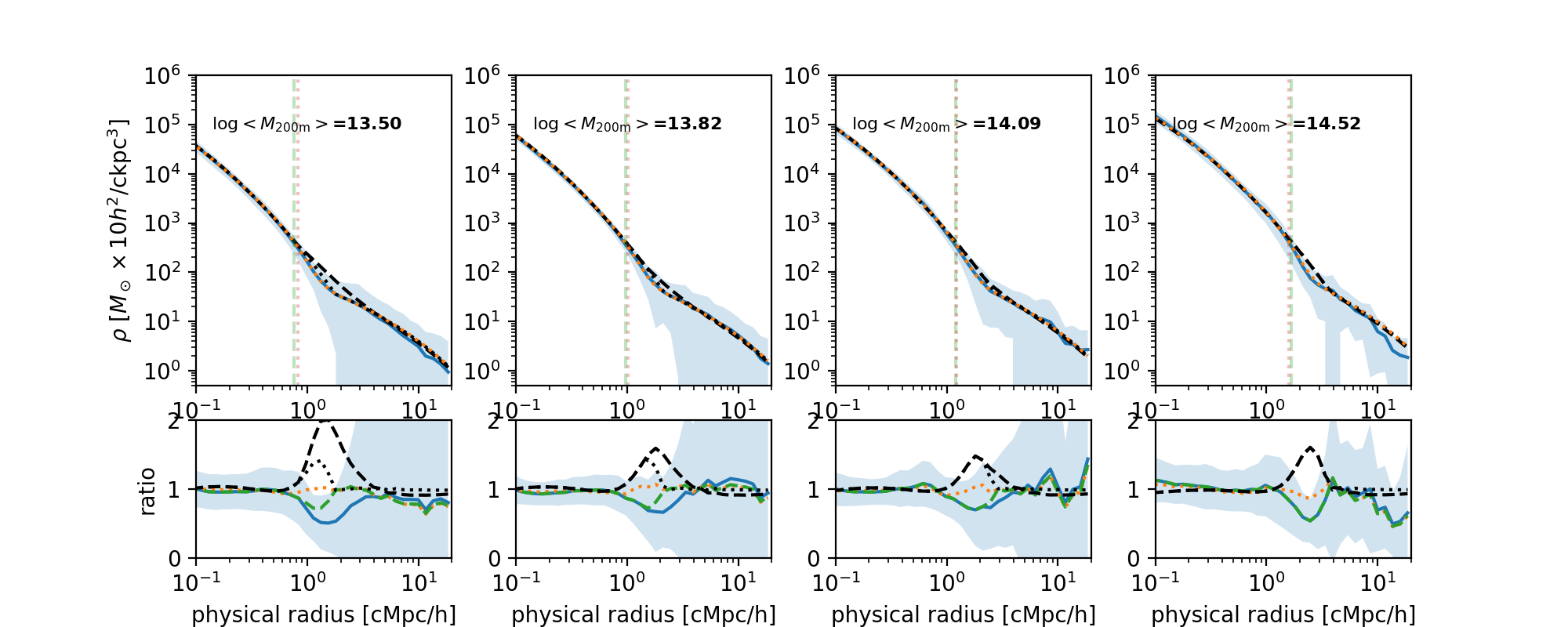}
	\includegraphics[width=2.5\columnwidth]{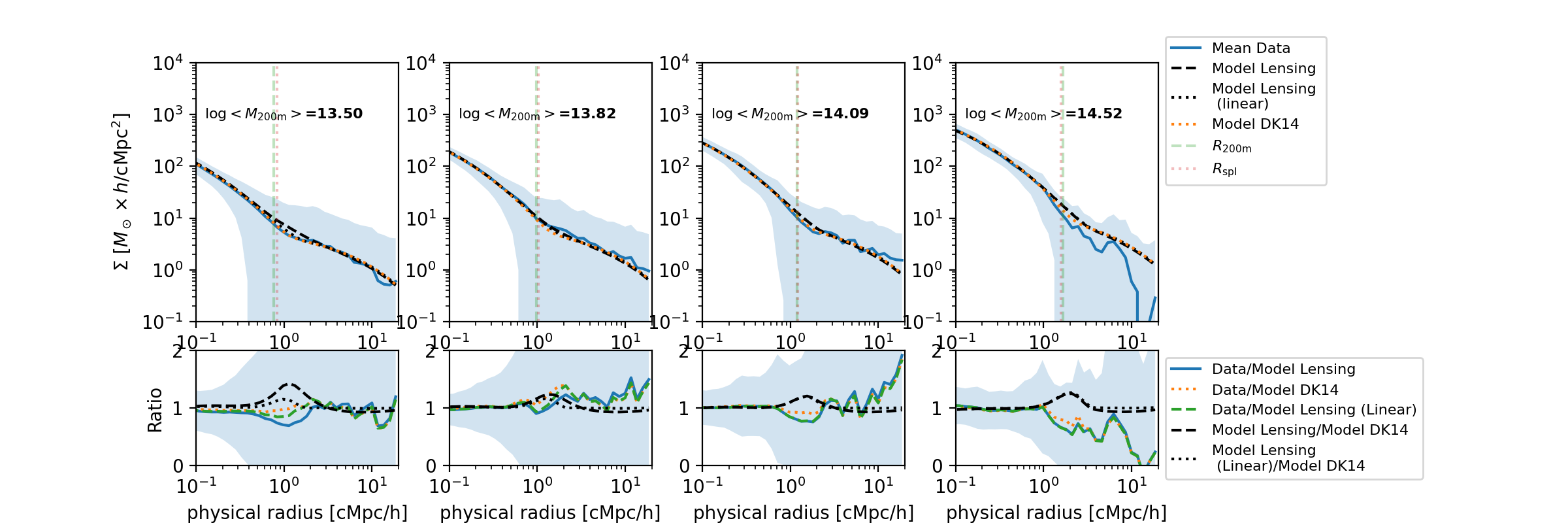}
	\includegraphics[width=2.1\columnwidth]{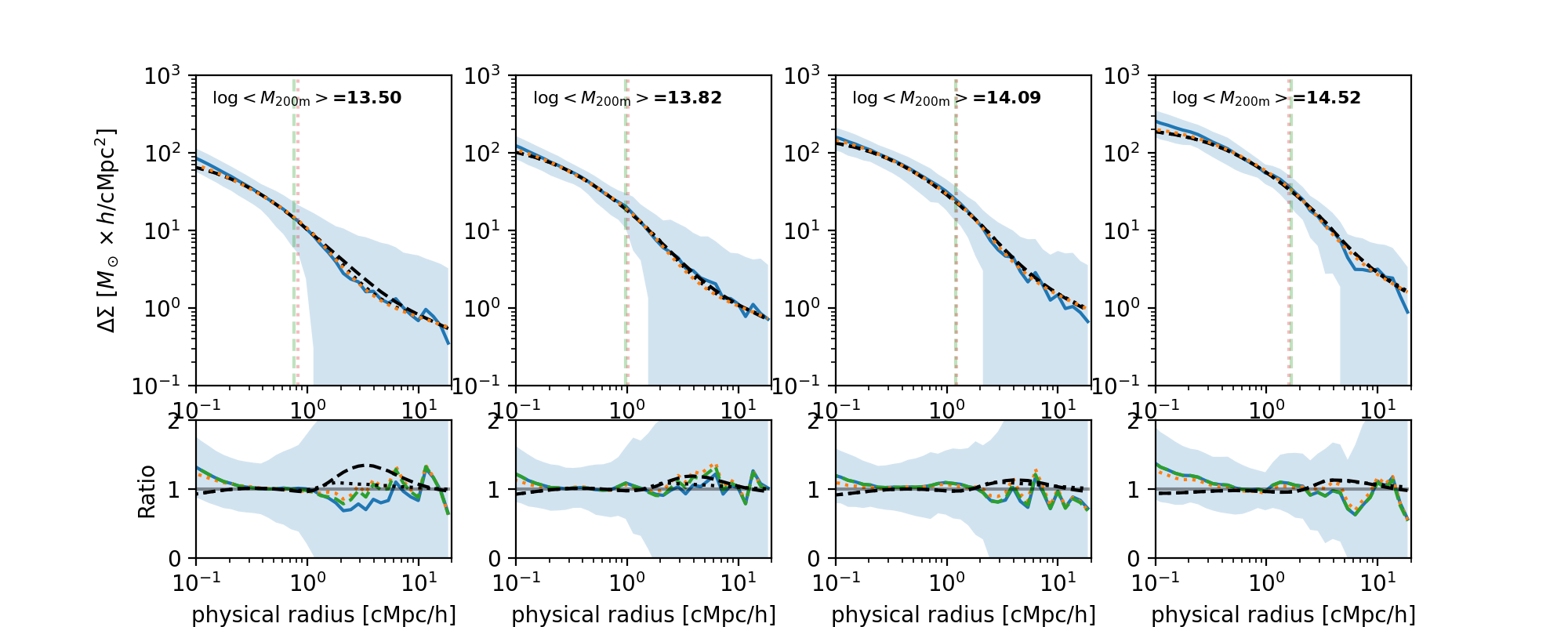}
    \caption{Average radial profiles and uncertainties (blue lines and blue shaded regions) of the cluster/group-sized dark matter halos in the TNG simulation in different mass ranges. From top to bottom: the 3D density profiles $\rho$, surface density profiles, $\Sigma$, and the differential surface density profiles $\Delta\Sigma$. From left to right, dark matter halos in different mass ranges, with average masses of $10^{13.5} \mathrm{M}_{\odot}$ on the left side and $10^{14.52} \mathrm{M}_{\odot}$ on the right side. The mean measurements of the profiles are shown as the solid blue lines in the larger panels, while the blue shaded regions indicate the $1\sigma$ halo-to-halo variations. We also show a lensing model (dashed black line), a lensing model with linear halo-matter correlation function based 2-halo term (dotted black line) and a splashback model (DK14, dotted orange line) in each panel and their differences to data (in the smaller panels). The splashback model better captures the splashback feature, especially for the low-mass halos.}
    \label{fig:covariance}
\end{figure*}

This analysis is based on the IllustrisTNG simulation \citep{2019ComAC...6....2N, 2018MNRAS.475..648P, 2018MNRAS.475..676S, 2018MNRAS.475..624N, 2018MNRAS.477.1206N, 2018MNRAS.480.5113M}. Specifically, we make use of the TNG 300-1 hydrodynamic simulation, which incorporates baryonic matter and has the highest resolution among the largest volume simulation sets in the IllustrisTNG suite. This simulation uses 2500$^3$ dark matter particles and an equal number of gas tracer particles in addition to  stellar particles within a volume of $205^3$ ($\mathrm{cMpc}/h)^3$ \footnote{$c$ in $\mathrm{cMpc}/h$ indicates comoving distance.}. The dark matter particle mass is $5.9\times 10^7 M_\odot$ while the average gas cell mass particle is $1.1\times 10^7 M_\odot$ which enables halo finding well below the group-sized halos studied here. This choice of simulation is motivated by the need to access high-resolution simulations with a large enough cosmic volume to analyze a few hundreds of dark matter halos in the galaxy cluster mass range. 

We analyze the dark matter halos with masses above $2\times 10^{13} \mathrm{M}_\odot /h$ in the redshift 0.27 snapshot. The purpose of this selection is to mimic the low-mass galaxy clusters probed by optical cosmic surveys like DES and the Sloan Digital Sky Survey \citep[SDSS][]{2010ApJ...708..645R, 2019MNRAS.488.4779C, 2021arXiv211209059P}. For each of those halos, we query the dark and baryonic matter particles in the simulation and compute the halo matter densities using the particles' information. When computing the 3D density profiles of the halos, we calculate each halo's matter density in spherical annulus using the particles' 3D distances to the halo density center. When computing the 2D projected density profiles, we project the particles onto the $X-Y$ plane of the simulation, using the particles' $X-Y$ distances to the halo density center on the $X-Y$ plane.

Because of large-scale structures, when analyzing the halo's projected densities, we need to select a projection depth that the profiles have converged at a percentage level. Following the analyses in \cite{2022MNRAS.511L..30Z}, we use a projection depth of $120~\mathrm{cMpc}/h$. We also use the same analysis strategy to only analyze half of the halo densities on the ``deep" side of the simulation to extend the available depth for projection: for each cluster-sized dark matter halo with a center coordinate of $(x_0,y_0, z_0)$, we only analyze one side of it along the $z-$axis that has a projection depth of $120~\mathrm{cMpc}/h$ available. The IllustrisTNG 300-1 simulation box has a length of $205~\mathrm{cMpc}/h$ in the $z$ direction. Therefore, for $z_0 < 85 ~\mathrm{cMpc}/h$, we only analyze the matter particles of $z> z_0$ in the simulation when deriving this halo's matter densities. If $z_0 > 120 ~\mathrm{cMpc}/h$, we only analyze the matter particles of $z< z_0$ in the simulation when deriving this halo's matter densities. The final derived matter densities are multiplied by 2 to recover the halo's full densities, assuming symmetry in the $z$ direction. Halos of $ 85 < z_0 < 120 ~\mathrm{cMpc}/h$ as well as those that are within 20 $\mathrm{cMpc}/h$ to any of the $X$, $Y$ and $Z$ boundaries are not analyzed because of insufficient projection depth or them being too close to the simulation box boundary. We have also tested projection depths of 75, 100 and 135 $\mathrm{cMpc}/h$, which would alter the projected halo matter densities presented in this paper. A smaller projection depth of 75 $\mathrm{cMpc}/h$ appears to reduce the projected matter densities by more than 10\% outside  1 $\mathrm{cMpc}/h$, possibly due to under-counting the contribution of large scale structures. On the other hand, using the 120 and 130 $\mathrm{cMpc}/h$ depths gives consistent results within 2\% in the 1 to 10 $\mathrm{cMpc}/h$, and the differences in the 10 to 20 $\mathrm{cMpc}/h$ radial range appear to be dominated by noise. Thus, we have opted to use the 120 $\mathrm{cMpc}/h$ depth in this analysis. 

The IllustrisTNG simulation has also enabled us to compare results between its hydrodynamic simulation suites and the dark-matter-only simulation suites. The splashback features discussed in this paper are present in both the hydrodynamic and dark-matter-only simulations, but the derived 3D density profiles differ by  $>5\%$ in the central 0 to 0.15 $\mathrm{cMpc}/h$ region of galaxy clusters above $10^{14.3} \mathrm{M}_\odot/h$, possibly due to baryonic effect. Anticipating better accuracy in the hydrodynamic simulations, we present results based on those simulations in this paper.

\section{The profiles and models}\label{sec:model}

\subsection{Radial Profiles}

Using the particle information from  the simulation, we proceed to derive the density profiles of the dark matter halos in different mass ranges. We compute their 3D density profiles ($\rho$) and the projected density profiles ($\Sigma$). To reduce noise in those measurements, those profiles are averaged for clusters in four $M_{200\mathrm{m}}$mass bins: $[2, 5]$, $[5, 9]$, $[9, 35]$ and $[35, \infty] \times 10^{13}\mathrm{M}_\odot/h$. The bin sizes are decided by visually checking the measurement uncertainty of $\rho$ to achieve a similar level of uncertainty in each bin. Further, the projected density profiles $\Sigma (r)$, are also converted into differential surface density profiles $\Delta \Sigma(r)$ as in 
\begin{equation} 
\Delta \Sigma(r)   =  \frac{2}{r^2}\int_0^r R \Sigma(R) \mathrm{d} R - \Sigma(r).
\end{equation}

The differential surface density  quantities are  more closely related to the measured tangential shear profile in weak lensing studies. The covariances for each of the $\rho(r)$, $\Sigma(r)$, or $\Delta\Sigma(r)$ profiles are computed as the halo-to-halo variation in each mass bin.

We compare the measurements from the simulation to the predictions from two different models. First, a one-halo and two-halo combination model adopted by a few cluster lensing analyses including \cite{2008MNRAS.388....2H, 2014MNRAS.439.1628Z, 2019MNRAS.482.1352M} that is based on the following halo-matter correlation function, 
\begin{equation}\label{eq:lensing}
    \begin{split}
        \xi(r|M, c) = \mathrm{max}\{\xi_\mathrm{NFW}(r|M, c), b(M)\xi_\mathrm{nl}(r)\}.
    \end{split}
\end{equation}
In this equation, $\xi_\mathrm{NFW}(r|M, c)$ is a halo-matter correlation function in the form of the NFW profile \citep{1996ApJ...462..563N} which depends on halo mass $M$ and concentration $c$, while $b(M) \xi_\mathrm{nl}(r)$ is a ``2-halo" term based on a non-linear halo-matter correlation function $\xi_\mathrm{nl}(r)$ \citep{2003MNRAS.341.1311S, 2012ApJ...761..152T} and a mass-dependent halo bias $b(M)$ \citep{2008ApJ...688..709T}. This halo-matter correlation function can be further converted into a 3D halo density function by multiplying it with the universe's mean density. We refer to this model as the ``lensing" model in the rest of this paper.

We also compare to a profile model that incorporates the splashback effect as introduced in \citet[][referred to as DK14 in the rest of this manuscript]{2014ApJ...789....1D}. The 3D density profile is written as 
\begin{equation}
    \begin{split}
        \rho(r|M, c) = \rho_\mathrm{Einasto}(r|M, c) \times f_\mathrm{trans}(r|M, c) + \rho(r|M).
    \end{split}
\end{equation}
This profile is based on an Einasto \citep{1969Ap......5...67E} profile $\rho_\mathrm{Einasto}(r|M, c)$ in the center modulated by a transition function $f_\mathrm{trans}(r|M, c)$, and converts into an outer profile at large radius which is based on the mass-dependent halo bias \citep{2008ApJ...688..709T} and a linear halo-matter correlation function \cite[][Appendix A2]{2014ApJ...789....1D}. We refer to this model as the ``splashback" model in the rest of this paper. Notably, the form of the ``splashback model" has been calibrated using N-body simulations for a wide halo mass range, adjusting for halo accretion rate and the halo outskirt/accretion region. Thus,  the ``splashback" model can better describe the splashback feature at the halo outskirt.

To compute those model profiles, we use the mean halo mass of each mass bin, and a concentration fitted in the later section ($\rho$ 0.2 to 1 cMpc$/h$ fitting) and then look up the profiles according to those mean masses and concentrations. 
Fig.~\ref{fig:covariance} shows the measurements  as well as the models. The relative differences between those models and the measurements are also displayed, and their $\chi^2$ values are listed in Table~\ref{tbl:chi2}.

Overall, at the high mass end when the halo mass is above $10^{14} \mathrm{M}_{\odot}/h$, the cluster density profiles appear to be well described by both the lensing model and the splashback model. However, below  $10^{14} \mathrm{M}_{\odot}/h$, the lensing model starts to display a noticeable discrepancy from the measurements around the splashback radius, and the discrepancy increases as the halo mass decreases. Its corresponding $\chi^2$ value also increases with decreasing halo mass, except in the highest mass bin. By contrast, the splashback models do not display the discrepancy around the splashback radius, indicating that the discrepancy in the lensing model is caused by the splashback features.

In addition, the discrepant radial ranges appear to change with  projections. We define a radial range of conceivable discrepancy when the difference between the lensing and splashback models is more than 20 \%. In the 3D density profiles, the discrepancy range is  between 0.80 to 3.0 cMpc$/h$ for the lowest halo mass bin. Switching to the projected density profiles, the splashback radial range moved inward because of the 3D to 2D projection, and the discrepancy range becomes 0.64 to 1.86 cMpc$/h$. Finally, the differential surface density profile used in lensing studies, 
 involves radial integration and differentiation of the projected densities. As a result, the differential surface density measurement not only further widens the affected radial range but also pushes the affected radial range outwards. The affected radial range for $\Delta\Sigma$ becomes 1.73 to 6.74 cMpc$/h$ for the lowest halo mass bin.

Notably, \cite{2017MNRAS.469.4899M} have also discussed a discrepancy between the lensing model and the halo matter densities in N-body simulations in the 1-halo to 2-halo transition region (Figure 8 of their paper), which is most significant in the lowest mass range of the dark matter halos in that analysis. The observational trend is consistent with our observations here, although the lowest mass bin in \cite{2017MNRAS.469.4899M} still has a higher mass range than the lowest mass bin in this analysis. A similar trend is also visible in \cite{2014MNRAS.439.1628Z}.

 Finally, when adopting a linear halo-matter correlation function in the ``2-halo" term in Equation~\ref{eq:lensing}, rather than the non-linear correlation function $\xi_\mathrm{nl}({r})$, appears to lessen the tension at the low mass end, which is also shown in Figure~\ref{fig:covariance}. This linear function-based lensing model appears to perform much better than the non-linear function based lensing model at the low-mass end, while being very similar to the latter at the high-mass end. However, it still performs slightly worse than the splashback model in the splashback radial range. Previously, this linear model was shown to be less accurate for cluster-sized halos in \cite{2014MNRAS.439.1628Z} in the  2-halo regime. For the rest of the work, we focus on presenting the lensing model with a 2-halo term based on the non-linear halo-matter correlation function, but note that switching to a linear function may be an acceptable solution to the mass-dependent bias investigated in this paper.

\subsection{Mass Bias}\label{sec:mass_bias}

\begin{figure}
	\includegraphics[width=1.0\columnwidth]{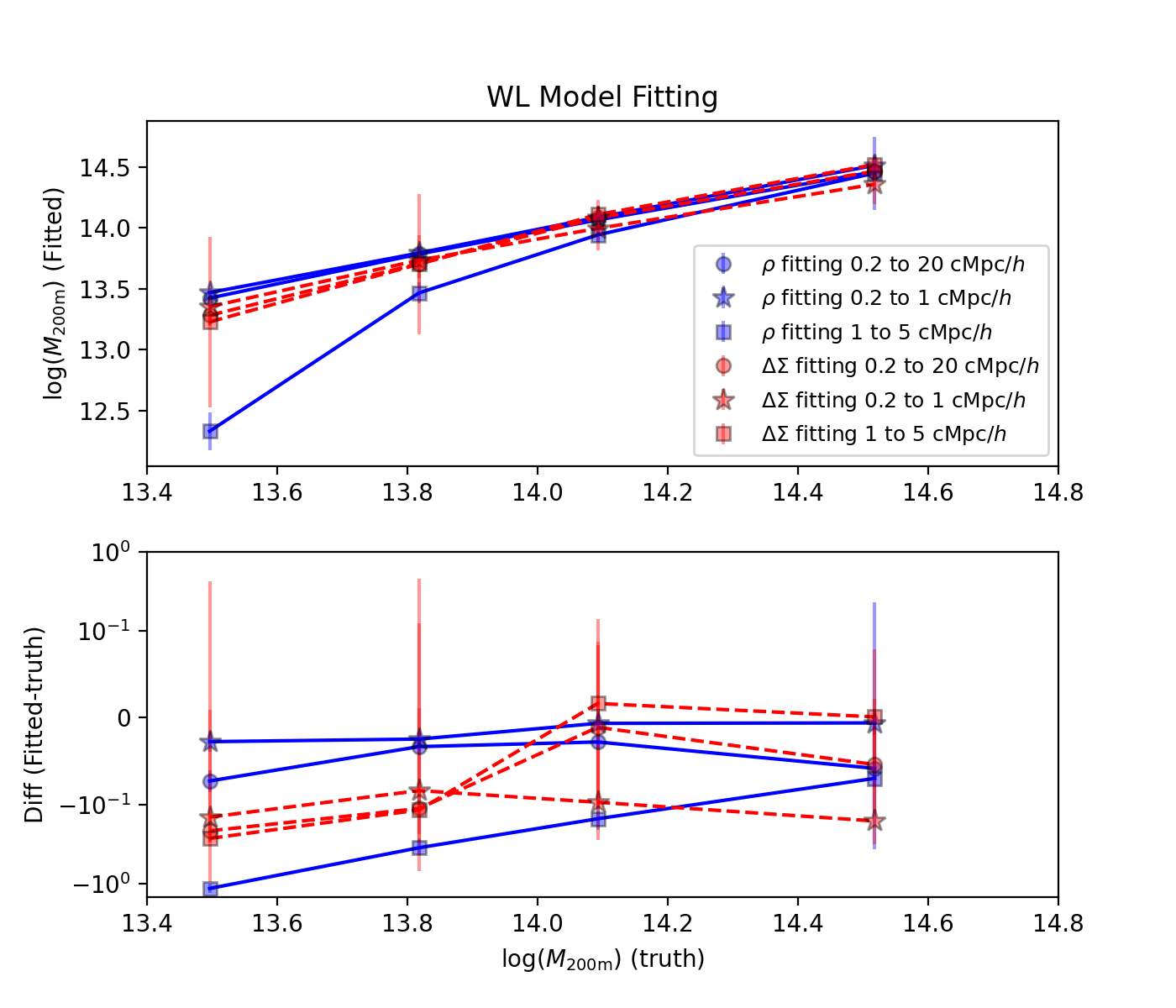}
	\includegraphics[width=1.0\columnwidth]{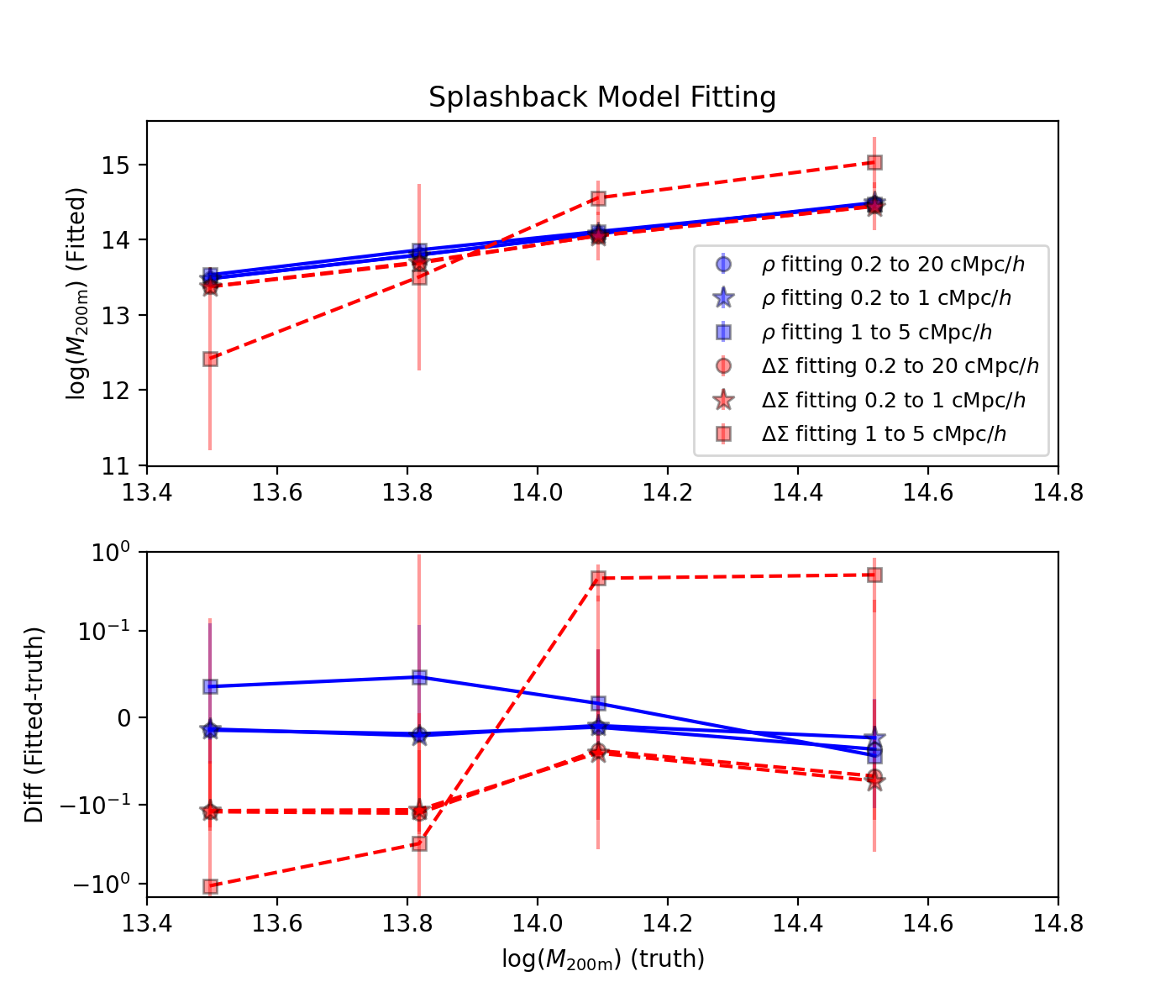}
    \caption{This figure shows the derived masses when fitting a lensing model (1st panel) and a splashback model (3rd panel) to the density profiles described in Figure 1 in different radial ranges. The biases of the fitted results are also shown (2nd and 4th panels, respectively, for the lensing and the splashback models), shown as the differences between the log values of the fitted masses and their truth values. For the lensing model, we note increasing mass bias towards the lower mass end, especially when fitting to radial ranges or the differential surface mass densities that are more susceptible to the splashback effect. On the other hand, using a splashback model significantly reduces the mass dependence of the bias. Note that we use a ``symlog" scale on the $y$-axis for the bias panels, that the $y$-axis scale is linear within $[-0.1, 0.1]$ but switches to being logarithmic outside that range. }
    \label{fig:mass_bias}
\end{figure}

\begin{table*}
\begin{tabular}{|l|lllll|} 
\hline
               & \multicolumn{1}{l|}{Model}         & \multicolumn{1}{l|}{$\mathrm{log}<M_\mathrm{200m}>$ $\sim$ 13.50} & \multicolumn{1}{l|}{$\mathrm{log}<M_\mathrm{200m}> \sim$13.82} & \multicolumn{1}{l|}{$\mathrm{log}<M_\mathrm{200m}> \sim$ 14.16} & $\mathrm{log}<M_\mathrm{200m}> \sim$ 14.52 \\ \hline
\rowcolor[HTML]{EFEFEF} 
$\rho$         & \multicolumn{5}{l|}{\cellcolor[HTML]{EFEFEF}}                                                                                                                                                                                                                                          \\ \hline
               & \multicolumn{1}{l|}{Lensing Model} & \multicolumn{1}{l|}{6.76}                                        & \multicolumn{1}{l|}{3.18}                                     & \multicolumn{1}{l|}{2.52}                                       & 7.21                                      \\ \hline
               & \multicolumn{1}{l|}{DK14 Model}    & \multicolumn{1}{l|}{0.40}                                         & \multicolumn{1}{l|}{0.61}                                      & \multicolumn{1}{l|}{0.55}                                       & 3.87                                      \\ \hline
\rowcolor[HTML]{EFEFEF} 
$\Sigma$       & \multicolumn{5}{l|}{\cellcolor[HTML]{EFEFEF}}                                                                                                                                                                                                                                          \\ \hline
               & \multicolumn{1}{l|}{Lensing Model} & \multicolumn{1}{l|}{0.36}                                         & \multicolumn{1}{l|}{0.11}                                      & \multicolumn{1}{l|}{0.35}                                       & 2.96                                     \\ \hline
               & \multicolumn{1}{l|}{DK14 Model}    & \multicolumn{1}{l|}{0.07}                                         & \multicolumn{1}{l|}{0.15}                                      & \multicolumn{1}{l|}{0.19}                                       & 2.50                                     \\ \hline
\rowcolor[HTML]{EFEFEF} 
$\Delta\Sigma$ & \multicolumn{5}{l|}{\cellcolor[HTML]{EFEFEF}}                                                                                                                                                                                                                                          \\ \hline
               & \multicolumn{1}{l|}{Lensing Model} & \multicolumn{1}{l|}{1.12}                                       & \multicolumn{1}{l|}{0.51}                                    & \multicolumn{1}{l|}{0.82}                                     & 2.52                                      \\ \hline
               & \multicolumn{1}{l|}{DK14 Model}    & \multicolumn{1}{l|}{0.69}                                         & \multicolumn{1}{l|}{0.26}                                      & \multicolumn{1}{l|}{0.30}                                       & 1.48                                      \\ \hline
\end{tabular}

\caption{$\chi^2$ between simulation measurements and the two models shown in Figure 1. The splashback DK14 models have lower $\chi^2$ values than the Lensing models except in the highest mass bin. The degree of freedom is 34 (34 radial values).}\label{tbl:chi2}
\end{table*}

Given the visible discrepancies in the density profiles, we further investigate how the model choices affect the halo mass derivations when the models are fitted to the measurements. As shown in the previous section, the lensing model based on the average halo mass tends to over-estimate the density profiles in the splashback region, thus for a lensing model to ``fit" to the measurements, the ``fitted" model would be expected to have a lower mass. 

For this estimation, we fit the lensing and splashback models to the derived radial profiles by allowing the mass and concentration of those models to vary. The fitting procedure is done through Markov Chain Monte Carlo (MCMC) assuming a Gaussian likelihood as in \begin{equation}
    \mathrm{log}\mathcal{L} = -\frac{1}{2}\vec{D}^T Cov^{-1}\vec{D}
\end{equation}
Here,  $\vec{D}$ is the vector of  difference between a model profile $F(r|M, c)$ and the measurement profile $M(r)$  as in $\vec{D} = F(r|M, c) - M(c)$. $M$ and $c$ are the mass and concentration, treated as the free parameters during the MCMC sampling process, with wide priors (a uniform distribution between 10.0 and 16.0 for $\mathrm{log}M$ and between 0.1 and 50.0 for $c$). $Cov$ is the covariance matrix derived as the halo-to-halo variations in each mass bin for each corresponding profile. %The fitting procedure is implemented with the EMCEE package \citep{2013PASP..125..306F}. 

Fig.~\ref{fig:mass_bias} shows the derived halo masses based on the 3D density  measurements and the differential projected density measurements. We also explore different radial ranges in the fitting process. 

When fitting a lensing model to the cluster density profiles, we indeed notice a tendency to underestimate halo mass with an increasing bias towards the lower mass end. In the 3D density case (blue lines in the upper two panels), the derived mass bias is most noticeable when fitting the 1 to 5 $\mathrm{cMpc}/h$ radial range, which is expected given that the splashback effect occurs around this radial range. Conversely, the mass bias seems harder to avoid fitting the 2D projected profiles: the masses derived from the 1 to 5 $\mathrm{cMpc}/h$, 0.2 to 20 $\mathrm{cMpc}/h$  fitting ranges both show some level of increasing biases. This may be partially related to the splashback affecting a wider radial range of those $\Delta \Sigma$ measurements because of projection and radial integration, as shown in the previous section. The derived mass bias in the lowest halo mass range with a mean mass of $10^{13.5} \mathrm{M}_\odot/h$ reaches 0.21 dex when fitting to $\Delta\Sigma$ in the 0.2 to 20 $\mathrm{cMpc}/h$ range, but is only 0.054 dex for the highest mass bin with a mean mass of $10^{14.52} \mathrm{M}_\odot/h$.

However, using the splashback model seems to alleviate much of the bias. When fitting to the differential projected density profiles $\Delta \Sigma$ in the 0.2 to 20 $\mathrm{cMpc}/h$ range, the splashback model appears to slightly under-estimate the mass by 0.12 dex for the lowest mass bin. However, this bias remains relatively stable for the other mass bins in contrast with the lensing model. The mass bias stands at 0.07 dex for the highest mass bin. Unfortunately, the mass derivation from the 1 to 5 $\mathrm{cMpc}/h$ radial range seems to lack constraining power compared to using the lensing model, which may be due to this range being a transition region between the Einasto profile and an outer profile component \citep[][]{2014ApJ...789....1D}. 

\subsection{Massive Galaxy-Sized Halos}

\begin{figure}
	%\includegraphics[width=\columnwidth]{measurement_profile.png}
	%\includegraphics[width=2.2\columnwidth]{Figure_1.png}
	\includegraphics[width=1.0\columnwidth]{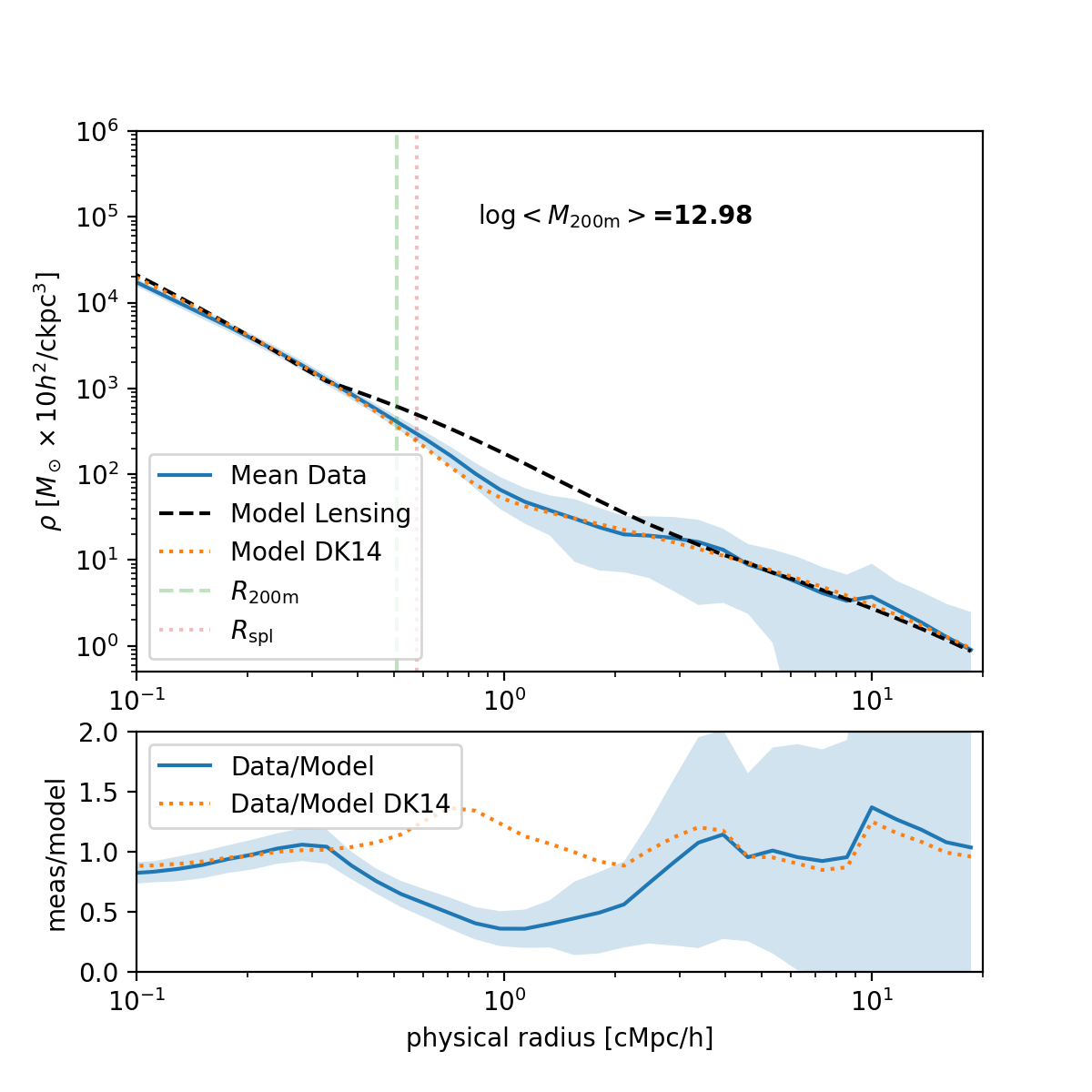}
    \caption{Average 3D radial profiles and uncertainties (blue lines and blue shaded regions) of the galaxy-sized dark matter halos in the mass range of $[0.9-1.0]\times 10^{13} \mathrm{M}_{\odot}$. We also show the lensing model (dashed blue line) and the splashback model (dotted orange line) for those dark matter halos, and their differences to the measurements (lower panel). Stark differences can be seen between the lensing model and the measurements. The splashback model appears to be more consistent. The profiles shown here are based on the halos' average mass and a concentration derived from the concentration-mass relation of \citet{2008MNRAS.390L..64D}. }%However, we recognize that halos in those mass ranges are analyzed differently from galaxy clusters, usually through large-scale cross-correlation analyses. Thus studies of those halos are not subject to the same cosmology and mass biases analyzed here.}
    \label{fig:low_mass}
\end{figure}

In the previous sections, we show that the splashback effect is increasingly significant at the low mass end for the cluster to group-sized dark matter halos. We further demonstrate the same effect for even lower-mass halos. 

Figure~\ref{fig:low_mass} shows the 3D density profiles of dark matter halos with masses between 0.9 to 1.0 $10^{13}\mathrm{M}_{\odot}/h$, which correspond to the hosts of massive galaxies. The deviation between the lensing model and the measurements remains stark, and the splashback model is still more consistent with the measurements. The same trend (not shown) holds for the 2D projected density profile and the differential surface density profile. 

We recognize that the lensing analysis of those galaxy-sized halos is different from the analysis of the cluster-sized halos; the former uses more complicated models for estimating large-scale structure contributions and is often estimated through galaxy-lensing or galaxy clustering analyses \citep[e.g., often known as ``3x2" analyses][]{2021arXiv210513548K}. Therefore, we do not make a mass bias estimation for those halos but do suggest the consideration of the splashback effect when analyzing their lensing signals.  

\section{Impacts on Cosmology}\label{sec:cosmo}

\subsection{Implication for a DES-like Analysis}

\begin{figure}
	%\includegraphics[width=\columnwidth]{measurement_profile.png}
	%\includegraphics[width=2.2\columnwidth]{Figure_1.png}
	\includegraphics[width=0.9\columnwidth]{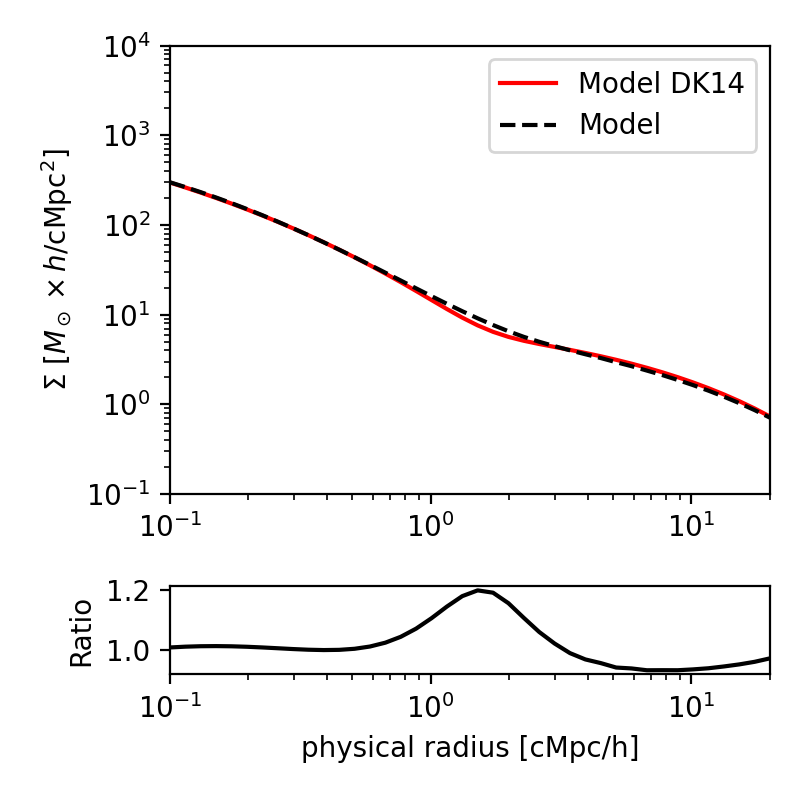}
    \caption{Reconstructed average density profiles of the dark matter halos studied in the lowest richness bin of DES at redshift 0.45, assuming a splashback model (red solid line) and a lensing model (black dashed line). The lensing model, as used in the DES analysis, predicts an excess of mass around the physical radius of 1 to 3 cMpc$/h$, but this excess is not as severe as that of the dark matter halos around $10^{13.5}M_\odot/h$. The splashback effect is not expected to explain the mass bias in the DES cluster analysis fully.}
    \label{fig:profile_bias}
\end{figure}

The number counts and the weak lensing signals of galaxy clusters are important inputs to cluster cosmological analyses. Therefore, we may expect a set of biased cosmological constraints if a discrepant model from the splashback effect is used in those analyses. Curiously, DES has discovered a bias that galaxy cluster mass $M_\mathrm{200m}$ appears to be under-estimated towards the lower-mass end in the weak lensing shear measurements, when compared to fixed cosmology expectation values. This mass bias, in turn, caused a bias in the derived cosmological constraints \citep{2020PhRvD.102b3509A}. The galaxy cluster masses in DES are measured from $\Delta\Sigma(r)$ measurements using the same lensing model studied in this paper. Simulation studies suggest that projection of large scale structures correlated with the galaxy cluster selection criterion based on galaxy overdensity may be the main cause of the DES mass bias \citep{2022MNRAS.515.4471W}.  %Since we are using the same lensing model with DES in this paper, we are concerned with whether or not the splashback effect had caused the bias. 

Although the splashback effect may produce a bias in a similar direction, it is unlikely to be the only source of the DES bias. %First, the DES analysis has adopted a calibration process that the lensing model is applied to a high-resolution dark matter simulation to test for potential mass bias.
First, DES has calibrated weak lensing mass systematics using a high-resolution dark matter simulation which should capture the splashback feature. The calibration process resulted in a  small  5 to 10\% correction term adopted in the DES analysis.  If the synthetic cluster catalog used in the calibration correctly reproduces the actual mass distribution of the observational cluster catalog, it will eliminate the mass bias caused by the splashback effect. Second, a study by \citet{2021arXiv211209059P} applied to similarly selected galaxy clusters from SDSS, but using an emulator-based lensing model (thus including the splashback features), also uncovered a similar cosmology discrepancy seen in the DES analysis.

Last but not the least, for the DES analysis, the low-mass dark matter halos only make up a small fraction of the clusters studied. Using the posterior constraints of cosmological parameters and the richness-mass scaling relation parameters, \citet{2020PhRvD.102b3509A} modeled the dark matter host mass distribution at redshift 0.45 in richness bins studied by DES. 
 In particular, even in the lowest richness bin\footnote{Richness is the mass proxy used in DES cluster cosmology analysis, which is based on a probabilistic count of cluster red-sequence galaxies.} considered in the analysis, only a small fraction of the clusters ($\sim12\%$) are below the halo mass threshold of  $10^{13.5}\mathrm{M}_{\odot}/h$ (see their Figure 9), which should result in a very small mass bias caused by the splashback effect, even if not calibrated.% even in the lowest richness bin considered in the analysis, only a small fraction of the clusters ($\sim12\%$) are below the halo mass threshold of  $10^{13.5}\mathrm{M}_{\odot}/h$, which should result in a small mass bias caused by the splashback effect, even if not calibrated. 

In Figure~\ref{fig:profile_bias},  we demonstrate the possible splashback effect in this lowest richness bin analyzed by DES -- the bin most likely affected by the splashback phenomenon. Using the halo mass distribution of this bin from \citet{2020PhRvD.102b3509A} (Figure 9), we compute a composite surface mass profile by averaging a mass model over that mass distribution $P(M)$ (which peaks around $10^{14.0}\mathrm{M}_{\odot}/h$), as in $\rho_\mathrm{composite} (r) = \int \mathrm{d} M P(M) \rho(r | M, c(M)) $, where, $\rho(r | M, c(M)) $ is either a lensing model or a splashback model, with concentration computed from a mass-concentration relation from \cite{2008MNRAS.390L..64D}. Compared to the lensing model, the composite profile based on the splashback  model has a lower surface mass density in the splashback radial region. However, throughout the 0.1 to 20 cMpc/$h$ radial range, the composite relative deficit is under 20\%, which is significantly lower than the difference previously seen in dark matter halos of $10^{13.5}\mathrm{M}_{\odot}/h$ (first column in Figure~\ref{fig:covariance}). Given that the mean mass of this richness bin is estimated to be around $10^{14.0}\mathrm{M}_{\odot}/h$ \citep[ranging from $10^{13.929}$ to $10^{14.036}$ $\mathrm{M}_{\odot}/h$ depending on the redshift range in][]{2020PhRvD.102b3509A, 2019MNRAS.482.1352M}, we can estimate a mass bias by interpolating the estimations in Section~\ref{sec:mass_bias} (Figure 2, $\Delta\Sigma$ fitting to 0.1 to 20 cMpc/$h$) at $10^{14.0}\mathrm{M}_{\odot}/h$, which yields an under-estimation bias of 0.09 dex, only 0.04 dex lower than the highest mass bin. The actual bias in the DES analysis caused by the splashback effect is likely much lower, given the additional mass calibration applied.

\subsection{Maximum Shift in the $\Omega_m$ Parameter}%Cosmology shift}

\begin{figure}
	%\includegraphics[width=\columnwidth]{measurement_profile.png}
	%\includegraphics[width=2.2\columnwidth]{Figure_1.png}
	\includegraphics[width=1.0\columnwidth]{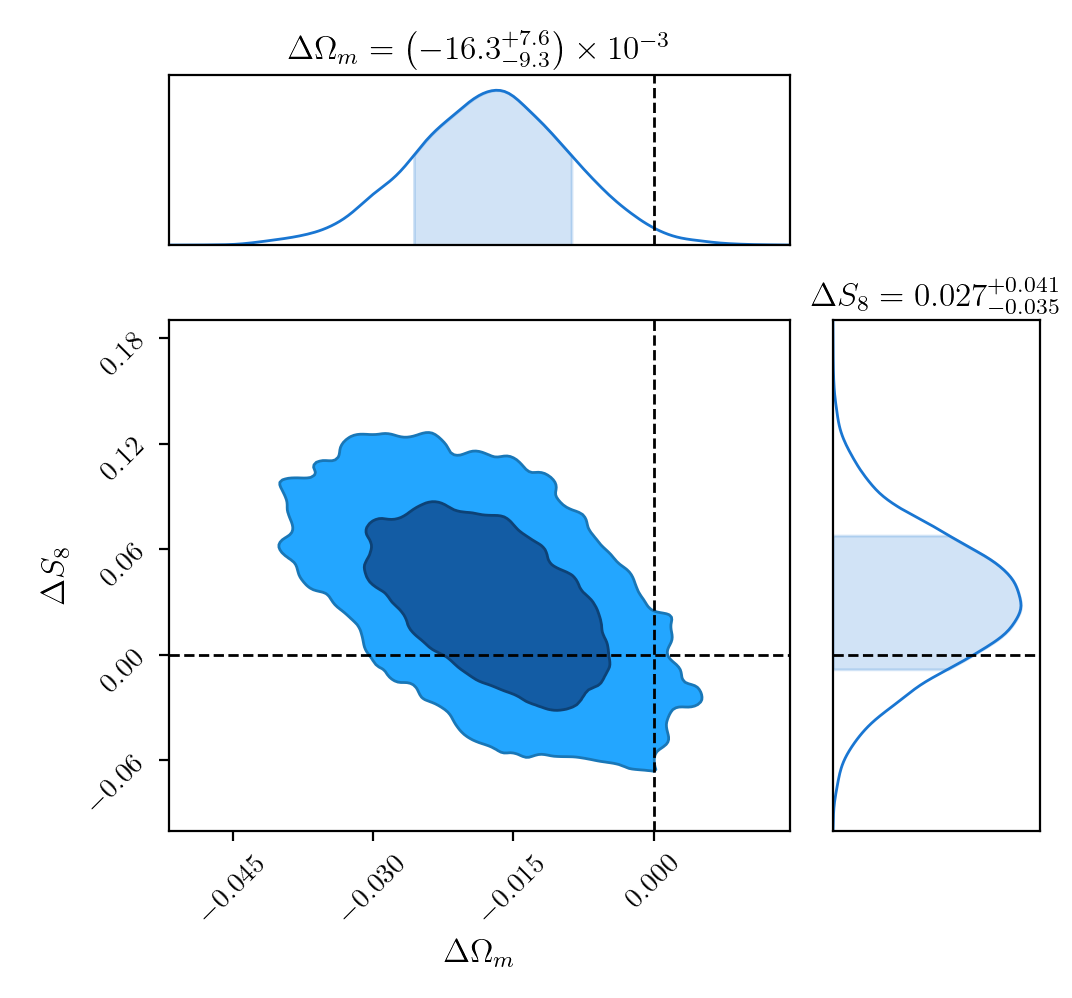}
    \caption{We derive $\Omega_m$ and $S_8$ (Defined as $\sigma_8 (\Omega_\mathrm{m}/0.3)^{0.5}$)from the IllustrisTNG 300-1 simulation at redshift 0.27 by modeling the dark matter halo counts above the mass cut of $2\times10^{13} \mathrm{M}_\odot/h$ and their 3D density profiles. This figure shows the cosmological shift between using the lensing model and using the splashback DK14 model, $\Delta \Omega_m$(WL - DK14) and $\Delta S_8$ (WL - DK14). This exercise is meant to explore the maximum cosmological impact in an idealized setup when including all low-mass dark matter halos above a mass threshold. We find that adopting a lensing model, which in previous sections has been shown to produce mass-dependent biases, yields lower $\Omega_m$ values than a splashback DK14 model.}
    \label{fig:cosmo}
\end{figure}

%Nevertheless, we want to demonstrate the impact on cosmology if the splashback effect is not sufficiently modeled for cluster/group-sized dark matter halos.
We have demonstrated that the splashback effect would generate a mass-dependent bias if not accounted for in the lensing model; it remains intriguing what the net effect would be on cosmological parameters if the bias is fully propagated into  cosmology. 

It is difficult to replicate a galaxy cluster cosmology analysis here, given the various elements of those analyses, including mass calibration and systematic effects. %Here, we demonstrate the cosmological effect with an idealized situation. 
Here we wish to qualitatively assess the maximum effect on the cosmological posteriors when incorrectly modeling the splashback feature. Using the dark matter halos above $2\times 10^{13} \mathrm{M}_\odot /h$ in the TNG simulation described in Section~\ref{sec:sims}, and assuming a $\Lambda$CDM cosmology, we derive the cosmological parameters through modeling the dark matter halo counts and their 3D matter density profiles. Specifically, we model the dark matter halo counts from the halo mass function. The models are written as the follows:
\begin{equation}
\begin{split}
  \left<N(\Delta M,z)\right> =  
    & V(z) \times  \int_{\Delta M} n(M,z) \mathrm{d}M \, \\
    \left<N \rho(r| \Delta M,z, c)\right> = & V(z) \times  \int_{\Delta M}  n(M,z) \rho_\mathrm{model}(r|M, z, c) \mathrm{d}M,  \\
    \left<\rho(r| \Delta M,z, c)\right> = & \left<N \rho(r| \Delta M,z, c)\right>/ \left<N(\Delta M,z)\right>
\end{split}
\label{eq:1}
\end{equation}
Here, $\left<N(\Delta M,z)\right>$ is the number of halos in a mass range of $\Delta M$ at redshift $z$. $\left<N \rho(r| \Delta M,z, c)\right>$ is the sum of their 3D matter density profiles as a function of $r$, in the mass range of $\Delta M$ at redshift $z$, assuming a concentration parameter of $c$. $\left<\rho(r| \Delta M,z, c)\right>$ is their averaged 3D matter density profiles. 

In those equations, $V(z)$ is the volume of the simulation, which is $205~c\mathrm{Mpc}/h^3$ (Note that this is a much smaller volume than observed by DES). $n(M, z)$ is the \citet[][]{2008ApJ...688..709T} halo mass function which can be analytically computed from the corresponding power spectrum of the cosmology. In computing the power spectrum, we adopt a $\Lambda$CDM cosmology with $\Omega_m$ and $\sigma_8$ as the varying parameters, while the rest of the parameters, like the Hubble parameter, having their values fixed at the simulation's truth. $\rho_\mathrm{model}(r|M, z, c)$ is the analytical model for the halo density profiles -- either the lensing model or the splashback model described in the previous section -- and $c$ is their concentration parameter. 

Note that the models and the observables adopted here are highly idealized compared to galaxy cluster observations: we assume that the dark matter halos are selected by their masses and redshifts directly, while in optical observations, galaxy clusters are selected by a mass ``proxy" such as richness and additional richness-mass scatters are introduced. Moreover,  we model the dark matter halos' 3D mass densities, which are not subject to projection effect -- another leading source of suspicion for causing cosmological biases.

To derive the best-fit cosmological parameters, we compare the model values to the measurements from the simulation using a likelihood function
\begin{equation}\label{eq:likelihood} 
\mathcal{L} = \mathcal{N}(x-\mu(\theta), C),
\end{equation}
with $x$ being the corresponding measurements in simulation, and $\mu(\theta)$ being the theoretical model values -- a vector of $\left<N(\Delta M,z)\right>$ and $\left<\rho(r| \Delta M,z, c)\right>$ described above -- which are dependent on the varying parameters $c$, $\Omega_m$ and $\sigma_8$. $C$ is the covariance matrix, and in this case, we define it as the measurement uncertainties; the measurement uncertainties of the halo counts are assumed to be Poissonian while the halo density profile uncertainties are the halo-to-halo variations described in the previous section. Both uncertainties are likely under-estimations as they do not account for noises caused by large-scale structures, and thus the posterior uncertainties of the constrained parameters are likely to be under-estimated as well. 

With this modeling setup, we perform Markov Chain Monte Carlo to sample $\Omega_m$, $\sigma_8$ as well as $c$, using either the lensing model or the splashback model. Given that this analysis contains many more low-mass clusters and groups than the DES analysis, our results show the maximum cosmological shift we can expect from a deep optical survey. For this reason, and the ones mentioned above, here we will limit our considerations only to the direction of the cosmological bias.

In Figure~\ref{fig:cosmo}, we show the shifts in the posterior results of  $\Omega_m$ and $S_8$, $\Delta \Omega_m$. Interestingly, the lensing model and the splashback model yield consistent $S_8$ (Defined as $\sigma_8 (\Omega_\mathrm{m}/0.3)^{0.5}$), but there exists a shift in the acquired $\Omega_m$ values. The lensing model, which produces a mass-dependent bias when fitted to the halo mass profiles directly, yields lower $\Omega_m$ values. Notably, the direction of the shift in $\Omega_m$ is consistent with the cosmological bias (lower $\Omega_m$ value than expected) revealed by DES in its galaxy cluster analysis. We again stress that the difference between a lensing model and a splashback model is unlikely to be the only source of the DES bias, given the reasons listed in the previous subsection. 

\section{summary}\label{sec:concl}

As optical surveys have enabled lensing and cosmological analyses of galaxy clusters and groups with masses lower than $10^{14}\mathrm{M}_\odot/h$, we suggest increasing caution about modeling the splashback effect of those objects. 

Using the baryonic simulation from the IllustrisTNG suite, we report findings that (1) the splashback effect grows more prominent when measuring the radial densities of dark matter halos below $10^{14}\mathrm{M}_\odot/h$ in mass. (2) A popular model used in cluster lensing studies consisting of an NFW-based one-halo term and a bias-based two-halo term appears to be increasingly insufficient to account for the splashback effect below $M_{200m}$ of $10^{14}\mathrm{M}_\odot/h$. Note that, although the studied mass ranges vary, \cite{2017MNRAS.469.4899M} have noted a similar shift towards low-mass halos in another simulation, thus this trend is unlikely to be dependent on details of simulations. (3) If not accounted for, the insufficiency can produce a mass-dependent bias in a similar direction to an observational bias revealed by a DES cluster cosmology study. The mass bias will also affect cluster cosmological constraints,  producing a lower $\Omega_m$ value, again in a similar direction to an observational bias revealed by DES. (4) The splashback effect, however, is unlikely to explain the bias discovered by DES fully.  %Nevertheless, we suggest future cluster cosmology analysis to properly account for the splashback effect to eliminate the biases.

Those splashback-related biases, however, appear to be easily eliminated by using models that explicitly model the splashback effect, such as \cite{2014ApJ...789....1D}. Further, Using a linear matter correlation function in the 2-halo term may be an acceptable solution for alleviating the mass-dependance of the bias. It is also possible that emulator-based models can account for this effect fully. While this manuscript points out potential issues when not considering the splashback phenomenon of galaxy clusters and groups, we are cautiously optimistic that they can be eliminated with improved modeling in  cluster lensing and cosmological studies.

\section*{Data Availability and Acknowledgements}

We are very grateful to the anonymous referee for the helpful comments that improved the quality of the work.

The data underlying this article were accessed from the Illustris-TNG database. The derived data generated in this research will be shared on reasonable request to the corresponding author.

The IllustrisTNG simulations were undertaken with compute time awarded by the Gauss Centre for Supercomputing (GCS) under GCS Large-Scale Projects GCS-ILLU and GCS-DWAR on the GCS share of the supercomputer Hazel Hen at the High Performance Computing Center Stuttgart (HLRS), as well as on the machines of the Max Planck Computing and Data Facility (MPCDF) in Garching, Germany.

Results in this paper made use of many software packages, including  \texttt{Astroy} \citep{astropy:2013, astropy:2018, astropy:2022}, \texttt{Numpy} \citep{harris2020array}, \texttt{Scipy} \citep{2020SciPy-NMeth}, \texttt{CosmoSIS} \citep{2014ascl.soft09012Z, 2015A&C....12...45Z}, \texttt{Emcee} \citep{2013PASP..125..306F}, and \texttt{ChainConsumer} \citep{Hinton2016}.

\bibliography{example}

\begin{thebibliography}{}
\makeatletter
\relax
\def\mn@urlcharsother{\let\do\@makeother \do\$\do\&\do\#\do\^\do\_\do\%\do\~}
\def\mn@doi{\begingroup\mn@urlcharsother \@ifnextchar [ {\mn@doi@}
  {\mn@doi@[]}}
\def\mn@doi@[#1]#2{\def\@tempa{#1}\ifx\@tempa\@empty \href
  {http://dx.doi.org/#2} {doi:#2}\else \href {http://dx.doi.org/#2} {#1}\fi
  \endgroup}
\def\mn@eprint#1#2{\mn@eprint@#1:#2::\@nil}
\def\mn@eprint@arXiv#1{\href {http://arxiv.org/abs/#1} {{\tt arXiv:#1}}}
\def\mn@eprint@dblp#1{\href {http://dblp.uni-trier.de/rec/bibtex/#1.xml}
  {dblp:#1}}
\def\mn@eprint@#1:#2:#3:#4\@nil{\def\@tempa {#1}\def\@tempb {#2}\def\@tempc
  {#3}\ifx \@tempc \@empty \let \@tempc \@tempb \let \@tempb \@tempa \fi \ifx
  \@tempb \@empty \def\@tempb {arXiv}\fi \@ifundefined
  {mn@eprint@\@tempb}{\@tempb:\@tempc}{\expandafter \expandafter \csname
  mn@eprint@\@tempb\endcsname \expandafter{\@tempc}}}

\bibitem[\protect\citeauthoryear{{Abbott} et~al.,}{{Abbott}
  et~al.}{2020}]{2020PhRvD.102b3509A}
{Abbott} T.~M.~C.,  et~al., 2020, \mn@doi [\prd] {10.1103/PhysRevD.102.023509},
  \href {https://ui.adsabs.harvard.edu/abs/2020PhRvD.102b3509A} {102, 023509}

\bibitem[\protect\citeauthoryear{{Adhikari}, {Dalal}  \&
  {Chamberlain}}{{Adhikari} et~al.}{2014}]{2014JCAP...11..019A}
{Adhikari} S.,  {Dalal} N.,   {Chamberlain} R.~T.,  2014, \mn@doi [\jcap]
  {10.1088/1475-7516/2014/11/019}, \href
  {https://ui.adsabs.harvard.edu/abs/2014JCAP...11..019A} {2014, 019}

\bibitem[\protect\citeauthoryear{{Adhikari}, {Sakstein}, {Jain}, {Dalal}  \&
  {Li}}{{Adhikari} et~al.}{2018}]{2018JCAP...11..033A}
{Adhikari} S.,  {Sakstein} J.,  {Jain} B.,  {Dalal} N.,   {Li} B.,  2018,
  \mn@doi [\jcap] {10.1088/1475-7516/2018/11/033}, \href
  {https://ui.adsabs.harvard.edu/abs/2018JCAP...11..033A} {2018, 033}

\bibitem[\protect\citeauthoryear{{Adhikari} et~al.,}{{Adhikari}
  et~al.}{2021}]{2021ApJ...923...37A}
{Adhikari} S.,  et~al., 2021, \mn@doi [\apj] {10.3847/1538-4357/ac0bbc}, \href
  {https://ui.adsabs.harvard.edu/abs/2021ApJ...923...37A} {923, 37}

\bibitem[\protect\citeauthoryear{{Astropy Collaboration} et~al.,}{{Astropy
  Collaboration} et~al.}{2013}]{astropy:2013}
{Astropy Collaboration} et~al., 2013, \mn@doi [\aap]
  {10.1051/0004-6361/201322068}, \href
  {http://adsabs.harvard.edu/abs/2013A%26A...558A..33A} {558, A33}

\bibitem[\protect\citeauthoryear{{Astropy Collaboration} et~al.,}{{Astropy
  Collaboration} et~al.}{2018}]{astropy:2018}
{Astropy Collaboration} et~al., 2018, \mn@doi [\aj] {10.3847/1538-3881/aabc4f},
  \href {https://ui.adsabs.harvard.edu/abs/2018AJ....156..123A} {156, 123}

\bibitem[\protect\citeauthoryear{{Astropy Collaboration} et~al.,}{{Astropy
  Collaboration} et~al.}{2022}]{astropy:2022}
{Astropy Collaboration} et~al., 2022, \mn@doi [apj] {10.3847/1538-4357/ac7c74},
  \href {https://ui.adsabs.harvard.edu/abs/2022ApJ...935..167A} {935, 167}

\bibitem[\protect\citeauthoryear{{Banerjee}, {Adhikari}, {Dalal}, {More}  \&
  {Kravtsov}}{{Banerjee} et~al.}{2020}]{2020JCAP...02..024B}
{Banerjee} A.,  {Adhikari} S.,  {Dalal} N.,  {More} S.,   {Kravtsov} A.,  2020,
  \mn@doi [\jcap] {10.1088/1475-7516/2020/02/024}, \href
  {https://ui.adsabs.harvard.edu/abs/2020JCAP...02..024B} {2020, 024}

\bibitem[\protect\citeauthoryear{{Baxter} et~al.,}{{Baxter}
  et~al.}{2017}]{2017ApJ...841...18B}
{Baxter} E.,  et~al., 2017, \mn@doi [\apj] {10.3847/1538-4357/aa6ff0}, \href
  {https://ui.adsabs.harvard.edu/abs/2017ApJ...841...18B} {841, 18}

\bibitem[\protect\citeauthoryear{{Bianconi}, {Buscicchio}, {Smith}, {McGee},
  {Haines}, {Finoguenov}  \& {Babul}}{{Bianconi}
  et~al.}{2021}]{2021ApJ...911..136B}
{Bianconi} M.,  {Buscicchio} R.,  {Smith} G.~P.,  {McGee} S.~L.,  {Haines}
  C.~P.,  {Finoguenov} A.,   {Babul} A.,  2021, \mn@doi [\apj]
  {10.3847/1538-4357/abebd7}, \href
  {https://ui.adsabs.harvard.edu/abs/2021ApJ...911..136B} {911, 136}

\bibitem[\protect\citeauthoryear{{Busch} \& {White}}{{Busch} \&
  {White}}{2017}]{2017MNRAS.470.4767B}
{Busch} P.,  {White} S. D.~M.,  2017, \mn@doi [\mnras] {10.1093/mnras/stx1584},
  \href {https://ui.adsabs.harvard.edu/abs/2017MNRAS.470.4767B} {470, 4767}

\bibitem[\protect\citeauthoryear{{Chang} et~al.,}{{Chang}
  et~al.}{2018}]{2018ApJ...864...83C}
{Chang} C.,  et~al., 2018, \mn@doi [\apj] {10.3847/1538-4357/aad5e7}, \href
  {https://ui.adsabs.harvard.edu/abs/2018ApJ...864...83C} {864, 83}

\bibitem[\protect\citeauthoryear{{Contigiani}, {Hoekstra}  \&
  {Bah{\'e}}}{{Contigiani} et~al.}{2019}]{2019MNRAS.485..408C}
{Contigiani} O.,  {Hoekstra} H.,   {Bah{\'e}} Y.~M.,  2019, \mn@doi [\mnras]
  {10.1093/mnras/stz404}, \href
  {https://ui.adsabs.harvard.edu/abs/2019MNRAS.485..408C} {485, 408}

\bibitem[\protect\citeauthoryear{{Costanzi} et~al.,}{{Costanzi}
  et~al.}{2019}]{2019MNRAS.488.4779C}
{Costanzi} M.,  et~al., 2019, \mn@doi [\mnras] {10.1093/mnras/stz1949}, \href
  {https://ui.adsabs.harvard.edu/abs/2019MNRAS.488.4779C} {488, 4779}

\bibitem[\protect\citeauthoryear{{Deason} et~al.,}{{Deason}
  et~al.}{2021}]{2021MNRAS.500.4181D}
{Deason} A.~J.,  et~al., 2021, \mn@doi [\mnras] {10.1093/mnras/staa3590}, \href
  {https://ui.adsabs.harvard.edu/abs/2021MNRAS.500.4181D} {500, 4181}

\bibitem[\protect\citeauthoryear{{Diemer}}{{Diemer}}{2021}]{2021ApJ...909..112D}
{Diemer} B.,  2021, \mn@doi [\apj] {10.3847/1538-4357/abd947}, \href
  {https://ui.adsabs.harvard.edu/abs/2021ApJ...909..112D} {909, 112}

\bibitem[\protect\citeauthoryear{{Diemer} \& {Kravtsov}}{{Diemer} \&
  {Kravtsov}}{2014}]{2014ApJ...789....1D}
{Diemer} B.,  {Kravtsov} A.~V.,  2014, \mn@doi [\apj]
  {10.1088/0004-637X/789/1/1}, \href
  {https://ui.adsabs.harvard.edu/abs/2014ApJ...789....1D} {789, 1}

\bibitem[\protect\citeauthoryear{{Diemer}, {Mansfield}, {Kravtsov}  \&
  {More}}{{Diemer} et~al.}{2017}]{2017ApJ...843..140D}
{Diemer} B.,  {Mansfield} P.,  {Kravtsov} A.~V.,   {More} S.,  2017, \mn@doi
  [\apj] {10.3847/1538-4357/aa79ab}, \href
  {https://ui.adsabs.harvard.edu/abs/2017ApJ...843..140D} {843, 140}

\bibitem[\protect\citeauthoryear{{Duffy}, {Schaye}, {Kay}  \& {Dalla
  Vecchia}}{{Duffy} et~al.}{2008}]{2008MNRAS.390L..64D}
{Duffy} A.~R.,  {Schaye} J.,  {Kay} S.~T.,   {Dalla Vecchia} C.,  2008, \mn@doi
  [\mnras] {10.1111/j.1745-3933.2008.00537.x}, \href
  {https://ui.adsabs.harvard.edu/abs/2008MNRAS.390L..64D} {390, L64}

\bibitem[\protect\citeauthoryear{{Einasto}}{{Einasto}}{1969}]{1969Ap......5...67E}
{Einasto} J.,  1969, \mn@doi [Astrophysics] {10.1007/BF01013353}, \href
  {https://ui.adsabs.harvard.edu/abs/1969Ap......5...67E} {5, 67}

\bibitem[\protect\citeauthoryear{{Foreman-Mackey}, {Hogg}, {Lang}  \&
  {Goodman}}{{Foreman-Mackey} et~al.}{2013}]{2013PASP..125..306F}
{Foreman-Mackey} D.,  {Hogg} D.~W.,  {Lang} D.,   {Goodman} J.,  2013, \mn@doi
  [\pasp] {10.1086/670067}, \href
  {https://ui.adsabs.harvard.edu/abs/2013PASP..125..306F} {125, 306}

\bibitem[\protect\citeauthoryear{{Gonzalez}, {George}, {Connor}, {Deason},
  {Donahue}, {Montes}, {Zabludoff}  \& {Zaritsky}}{{Gonzalez}
  et~al.}{2021}]{2021MNRAS.507..963G}
{Gonzalez} A.~H.,  {George} T.,  {Connor} T.,  {Deason} A.,  {Donahue} M.,
  {Montes} M.,  {Zabludoff} A.~I.,   {Zaritsky} D.,  2021, \mn@doi [\mnras]
  {10.1093/mnras/stab2117}, \href
  {https://ui.adsabs.harvard.edu/abs/2021MNRAS.507..963G} {507, 963}

\bibitem[\protect\citeauthoryear{Harris et~al.,}{Harris
  et~al.}{2020}]{harris2020array}
Harris C.~R.,  et~al., 2020, \mn@doi [Nature] {10.1038/s41586-020-2649-2}, 585,
  357

\bibitem[\protect\citeauthoryear{{Hayashi} \& {White}}{{Hayashi} \&
  {White}}{2008}]{2008MNRAS.388....2H}
{Hayashi} E.,  {White} S. D.~M.,  2008, \mn@doi [\mnras]
  {10.1111/j.1365-2966.2008.13371.x}, \href
  {https://ui.adsabs.harvard.edu/abs/2008MNRAS.388....2H} {388, 2}

\bibitem[\protect\citeauthoryear{{Hinton}}{{Hinton}}{2016}]{Hinton2016}
{Hinton} S.~R.,  2016, \mn@doi [The Journal of Open Source Software]
  {10.21105/joss.00045}, \href
  {http://adsabs.harvard.edu/abs/2016JOSS....1...45H} {1, 00045}

\bibitem[\protect\citeauthoryear{{Krause} et~al.,}{{Krause}
  et~al.}{2021}]{2021arXiv210513548K}
{Krause} E.,  et~al., 2021, \mn@doi [arXiv e-prints]
  {10.48550/arXiv.2105.13548}, \href
  {https://ui.adsabs.harvard.edu/abs/2021arXiv210513548K} {p. arXiv:2105.13548}

\bibitem[\protect\citeauthoryear{{Mansfield}, {Kravtsov}  \&
  {Diemer}}{{Mansfield} et~al.}{2017}]{2017ApJ...841...34M}
{Mansfield} P.,  {Kravtsov} A.~V.,   {Diemer} B.,  2017, \mn@doi [\apj]
  {10.3847/1538-4357/aa7047}, \href
  {https://ui.adsabs.harvard.edu/abs/2017ApJ...841...34M} {841, 34}

\bibitem[\protect\citeauthoryear{{Marinacci} et~al.,}{{Marinacci}
  et~al.}{2018}]{2018MNRAS.480.5113M}
{Marinacci} F.,  et~al., 2018, \mn@doi [\mnras] {10.1093/mnras/sty2206}, \href
  {https://ui.adsabs.harvard.edu/abs/2018MNRAS.480.5113M} {480, 5113}

\bibitem[\protect\citeauthoryear{{McClintock} et~al.,}{{McClintock}
  et~al.}{2019}]{2019MNRAS.482.1352M}
{McClintock} T.,  et~al., 2019, \mn@doi [\mnras] {10.1093/mnras/sty2711}, \href
  {https://ui.adsabs.harvard.edu/abs/2019MNRAS.482.1352M} {482, 1352}

\bibitem[\protect\citeauthoryear{{Melchior} et~al.,}{{Melchior}
  et~al.}{2017}]{2017MNRAS.469.4899M}
{Melchior} P.,  et~al., 2017, \mn@doi [\mnras] {10.1093/mnras/stx1053}, \href
  {https://ui.adsabs.harvard.edu/abs/2017MNRAS.469.4899M} {469, 4899}

\bibitem[\protect\citeauthoryear{{Miyatake}, {More}, {Takada}, {Spergel},
  {Mandelbaum}, {Rykoff}  \& {Rozo}}{{Miyatake}
  et~al.}{2016}]{2016PhRvL.116d1301M}
{Miyatake} H.,  {More} S.,  {Takada} M.,  {Spergel} D.~N.,  {Mandelbaum} R.,
  {Rykoff} E.~S.,   {Rozo} E.,  2016, \mn@doi [\prl]
  {10.1103/PhysRevLett.116.041301}, \href
  {https://ui.adsabs.harvard.edu/abs/2016PhRvL.116d1301M} {116, 041301}

\bibitem[\protect\citeauthoryear{{Miyatake} et~al.,}{{Miyatake}
  et~al.}{2019}]{2019ApJ...875...63M}
{Miyatake} H.,  et~al., 2019, \mn@doi [\apj] {10.3847/1538-4357/ab0af0}, \href
  {https://ui.adsabs.harvard.edu/abs/2019ApJ...875...63M} {875, 63}

\bibitem[\protect\citeauthoryear{{More}, {Diemer}  \& {Kravtsov}}{{More}
  et~al.}{2015}]{2015ApJ...810...36M}
{More} S.,  {Diemer} B.,   {Kravtsov} A.~V.,  2015, \mn@doi [\apj]
  {10.1088/0004-637X/810/1/36}, \href
  {https://ui.adsabs.harvard.edu/abs/2015ApJ...810...36M} {810, 36}

\bibitem[\protect\citeauthoryear{{More} et~al.,}{{More}
  et~al.}{2016}]{2016ApJ...825...39M}
{More} S.,  et~al., 2016, \mn@doi [\apj] {10.3847/0004-637X/825/1/39}, \href
  {https://ui.adsabs.harvard.edu/abs/2016ApJ...825...39M} {825, 39}

\bibitem[\protect\citeauthoryear{{Naiman} et~al.,}{{Naiman}
  et~al.}{2018}]{2018MNRAS.477.1206N}
{Naiman} J.~P.,  et~al., 2018, \mn@doi [\mnras] {10.1093/mnras/sty618}, \href
  {https://ui.adsabs.harvard.edu/abs/2018MNRAS.477.1206N} {477, 1206}

\bibitem[\protect\citeauthoryear{{Navarro}, {Frenk}  \& {White}}{{Navarro}
  et~al.}{1996}]{1996ApJ...462..563N}
{Navarro} J.~F.,  {Frenk} C.~S.,   {White} S. D.~M.,  1996, \mn@doi [\apj]
  {10.1086/177173}, \href
  {https://ui.adsabs.harvard.edu/abs/1996ApJ...462..563N} {462, 563}

\bibitem[\protect\citeauthoryear{{Nelson} et~al.,}{{Nelson}
  et~al.}{2018}]{2018MNRAS.475..624N}
{Nelson} D.,  et~al., 2018, \mn@doi [\mnras] {10.1093/mnras/stx3040}, \href
  {https://ui.adsabs.harvard.edu/abs/2018MNRAS.475..624N} {475, 624}

\bibitem[\protect\citeauthoryear{{Nelson} et~al.,}{{Nelson}
  et~al.}{2019}]{2019ComAC...6....2N}
{Nelson} D.,  et~al., 2019, \mn@doi [Computational Astrophysics and Cosmology]
  {10.1186/s40668-019-0028-x}, \href
  {https://ui.adsabs.harvard.edu/abs/2019ComAC...6....2N} {6, 2}

\bibitem[\protect\citeauthoryear{{Park}, {Sunayama}, {Takada}, {Kobayashi},
  {Miyatake}, {More}, {Nishimichi}  \& {Sugiyama}}{{Park}
  et~al.}{2021}]{2021arXiv211209059P}
{Park} Y.,  {Sunayama} T.,  {Takada} M.,  {Kobayashi} Y.,  {Miyatake} H.,
  {More} S.,  {Nishimichi} T.,   {Sugiyama} S.,  2021, arXiv e-prints, \href
  {https://ui.adsabs.harvard.edu/abs/2021arXiv211209059P} {p. arXiv:2112.09059}

\bibitem[\protect\citeauthoryear{{Pillepich} et~al.,}{{Pillepich}
  et~al.}{2018}]{2018MNRAS.475..648P}
{Pillepich} A.,  et~al., 2018, \mn@doi [\mnras] {10.1093/mnras/stx3112}, \href
  {https://ui.adsabs.harvard.edu/abs/2018MNRAS.475..648P} {475, 648}

\bibitem[\protect\citeauthoryear{{Rozo} et~al.,}{{Rozo}
  et~al.}{2010}]{2010ApJ...708..645R}
{Rozo} E.,  et~al., 2010, \mn@doi [\apj] {10.1088/0004-637X/708/1/645}, \href
  {https://ui.adsabs.harvard.edu/abs/2010ApJ...708..645R} {708, 645}

\bibitem[\protect\citeauthoryear{Shi}{Shi}{2016}]{Shi:2016lwp}
Shi X.,  2016, \mn@doi [Mon. Not. Roy. Astron. Soc.] {10.1093/mnras/stw925},
  459, 3711

\bibitem[\protect\citeauthoryear{{Shin} et~al.,}{{Shin}
  et~al.}{2019}]{2019MNRAS.487.2900S}
{Shin} T.,  et~al., 2019, \mn@doi [\mnras] {10.1093/mnras/stz1434}, \href
  {https://ui.adsabs.harvard.edu/abs/2019MNRAS.487.2900S} {487, 2900}

\bibitem[\protect\citeauthoryear{{Smith} et~al.,}{{Smith}
  et~al.}{2003}]{2003MNRAS.341.1311S}
{Smith} R.~E.,  et~al., 2003, \mn@doi [\mnras]
  {10.1046/j.1365-8711.2003.06503.x}, \href
  {https://ui.adsabs.harvard.edu/abs/2003MNRAS.341.1311S} {341, 1311}

\bibitem[\protect\citeauthoryear{{Springel} et~al.,}{{Springel}
  et~al.}{2018}]{2018MNRAS.475..676S}
{Springel} V.,  et~al., 2018, \mn@doi [\mnras] {10.1093/mnras/stx3304}, \href
  {https://ui.adsabs.harvard.edu/abs/2018MNRAS.475..676S} {475, 676}

\bibitem[\protect\citeauthoryear{{Sunayama} \& {More}}{{Sunayama} \&
  {More}}{2019}]{2019MNRAS.490.4945S}
{Sunayama} T.,  {More} S.,  2019, \mn@doi [\mnras] {10.1093/mnras/stz2832},
  \href {https://ui.adsabs.harvard.edu/abs/2019MNRAS.490.4945S} {490, 4945}

\bibitem[\protect\citeauthoryear{{Takahashi}, {Sato}, {Nishimichi}, {Taruya}
  \& {Oguri}}{{Takahashi} et~al.}{2012}]{2012ApJ...761..152T}
{Takahashi} R.,  {Sato} M.,  {Nishimichi} T.,  {Taruya} A.,   {Oguri} M.,
  2012, \mn@doi [\apj] {10.1088/0004-637X/761/2/152}, \href
  {https://ui.adsabs.harvard.edu/abs/2012ApJ...761..152T} {761, 152}

\bibitem[\protect\citeauthoryear{{Tinker}, {Kravtsov}, {Klypin}, {Abazajian},
  {Warren}, {Yepes}, {Gottl{\"o}ber}  \& {Holz}}{{Tinker}
  et~al.}{2008}]{2008ApJ...688..709T}
{Tinker} J.,  {Kravtsov} A.~V.,  {Klypin} A.,  {Abazajian} K.,  {Warren} M.,
  {Yepes} G.,  {Gottl{\"o}ber} S.,   {Holz} D.~E.,  2008, \mn@doi [\apj]
  {10.1086/591439}, \href
  {https://ui.adsabs.harvard.edu/abs/2008ApJ...688..709T} {688, 709}

\bibitem[\protect\citeauthoryear{{Umetsu} \& {Diemer}}{{Umetsu} \&
  {Diemer}}{2017}]{2017ApJ...836..231U}
{Umetsu} K.,  {Diemer} B.,  2017, \mn@doi [\apj] {10.3847/1538-4357/aa5c90},
  \href {https://ui.adsabs.harvard.edu/abs/2017ApJ...836..231U} {836, 231}

\bibitem[\protect\citeauthoryear{Virtanen et~al.,}{Virtanen
  et~al.}{2020}]{2020SciPy-NMeth}
Virtanen P.,  et~al., 2020, \mn@doi [Nature Methods]
  {10.1038/s41592-019-0686-2}, \href {https://rdcu.be/b08Wh} {17, 261}

\bibitem[\protect\citeauthoryear{{Wu} et~al.,}{{Wu}
  et~al.}{2022}]{2022MNRAS.515.4471W}
{Wu} H.-Y.,  et~al., 2022, \mn@doi [\mnras] {10.1093/mnras/stac2048}, \href
  {https://ui.adsabs.harvard.edu/abs/2022MNRAS.515.4471W} {515, 4471}

\bibitem[\protect\citeauthoryear{{Zhang} \& {Annis}}{{Zhang} \&
  {Annis}}{2022}]{2022MNRAS.511L..30Z}
{Zhang} Y.,  {Annis} J.,  2022, \mn@doi [\mnras] {10.1093/mnrasl/slac002},
  \href {https://ui.adsabs.harvard.edu/abs/2022MNRAS.511L..30Z} {511, L30}

\bibitem[\protect\citeauthoryear{{Zu}, {Weinberg}, {Rozo}, {Sheldon}, {Tinker}
  \& {Becker}}{{Zu} et~al.}{2014}]{2014MNRAS.439.1628Z}
{Zu} Y.,  {Weinberg} D.~H.,  {Rozo} E.,  {Sheldon} E.~S.,  {Tinker} J.~L.,
  {Becker} M.~R.,  2014, \mn@doi [\mnras] {10.1093/mnras/stu033}, \href
  {https://ui.adsabs.harvard.edu/abs/2014MNRAS.439.1628Z} {439, 1628}

\bibitem[\protect\citeauthoryear{{Zuntz} et~al.,}{{Zuntz}
  et~al.}{2014}]{2014ascl.soft09012Z}
{Zuntz} J.,  et~al., 2014, {CosmoSIS: Cosmological parameter estimation},
  Astrophysics Source Code Library, record ascl:1409.012 (\mn@eprint {ascl}
  {1409.012})

\bibitem[\protect\citeauthoryear{{Zuntz} et~al.,}{{Zuntz}
  et~al.}{2015}]{2015A&C....12...45Z}
{Zuntz} J.,  et~al., 2015, \mn@doi [Astronomy and Computing]
  {10.1016/j.ascom.2015.05.005}, \href
  {https://ui.adsabs.harvard.edu/abs/2015A&C....12...45Z} {12, 45}

\makeatother
\end{thebibliography}

\end{document}